\documentclass[10pt,journal,compsoc]{IEEEtran}
\ifCLASSOPTIONcompsoc
  \usepackage[nocompress]{cite}
\else
  \usepackage{cite}
\fi
\ifCLASSINFOpdf
\else
\fi
\usepackage{epsfig}
\usepackage{comment}
\usepackage[pdftex]{color}
\usepackage{url}
\usepackage{listings}
\usepackage{textcomp}
\usepackage{amssymb}
\usepackage{amsmath}
\usepackage{paralist}
\usepackage{subcaption}
\usepackage{algorithm}
\usepackage[noend]{algpseudocode}
\usepackage{graphicx}
\usepackage{float}
\usepackage{authblk}
\usepackage{epstopdf}
\usepackage{soul}
\definecolor{lbcolor}{rgb}{0.9,0.9,0.9}

\begin{document}
%
\title{Semi-External Memory Sparse Matrix Multiplication for Billion-Node Graphs} 
%
%
%
%

\author[1]{\rm Da Zheng}
\author[1]{\rm Disa Mhembere}
\author[2]{\rm Vince Lyzinski}
\author[3]{\rm Joshua T. Vogelstein}
\author[2]{\rm Carey E. Priebe}
\author[1]{\rm Randal Burns}
\affil[1]{Department of Computer Science, Johns Hopkins University}
\affil[2]{Department of Applied Mathematics and Statistics, Johns Hopkins University}
\affil[3]{Department of Biomedical Engineering, Johns Hopkins University}

\IEEEtitleabstractindextext{%
\begin{abstract}
%
Sparse matrix multiplication is traditionally performed in memory and scales to large matrices using
the distributed memory of multiple nodes.
In contrast, we scale sparse matrix multiplication beyond memory capacity 
by implementing sparse matrix dense matrix multiplication (SpMM) in
a semi-external memory (SEM) fashion; i.e., we keep the sparse matrix on commodity SSDs
and dense matrices in memory. Our SEM-SpMM incorporates many
in-memory optimizations for large power-law graphs.
It outperforms the in-memory implementations of Trilinos
and Intel MKL and scales to billion-node graphs, far beyond the limitations of memory.
Furthermore, on a single large parallel machine, our SEM-SpMM operates as fast as
the distributed implementations of Trilinos using five times as much processing power.
We also run our implementation in memory (IM-SpMM) to quantify the overhead of
keeping data on SSDs.  SEM-SpMM achieves almost 100\% performance of IM-SpMM
on graphs when the dense matrix has more than four columns; it achieves at least
65\% performance of IM-SpMM on all inputs. 
We apply our SpMM to three important data analysis tasks---PageRank, eigensolving, and 
non-negative matrix factorization---and show that
our SEM implementations significantly advance the state of the art.
\end{abstract}

}

\maketitle

\IEEEdisplaynontitleabstractindextext

%
\IEEEpeerreviewmaketitle

\section{Introduction}
Sparse matrix multiplication is an important computation with a wide variety
of applications in scientific computing, machine learning, and graph analysis.
For example, matrix factorization algorithms on a sparse matrix, such as
singular value decomposition (SVD) \cite{svd} and non-negative matrix
factorization (NMF) \cite{nmf}, require sparse matrix multiplication.
Graph analysis algorithms such as PageRank \cite{pagerank} can be
formulated as sparse matrix multiplication or generalized sparse matrix
multiplication \cite{Mattson13}. Some of
the algorithms, such as PageRank and SVD, require sparse matrix vector
multiplication. Others, such as NMF, require sparse matrix dense
matrix multiplication.

It is challenging to implement sparse matrix multiplication efficiently. 
It has very low computation density and its
performance is limited by memory access. Sparse matrix multiplication usually achieves only
a small fraction of the peak performance of a modern processor \cite{Williams07}.

Sparse matrices derived from graphs impose additional challenges.
Many graphs have a power-law distribution in vertex
degree.  This produces a power-law distribution in the number of non-zero 
entries per column or row in the associated matrix, which   
results in load imbalance in parallel decompositions of sparse 
matrix multiplication. 
Graph matrices are much sparser and have more random distribution 
than traditional sparse matrices that arise in scientific
computing. As such, many optimizations for traditional sparse matrices, such
as register blocking and prefetching \cite{Williams07} increase storage size,
bring more data to CPU cache unnecessarily and hurt performance.
The commonly used sparse matrix storage formats such as CSR (compressed sparse
row) and CSC (compressed sparse column) for graph matrices leads to significant
CPU cache misses in sparse matrix multiplication because graphs are sparse and
large.

We focus on sparse matrices derived from graphs because they are more difficult 
to compute efficiently than general sparse matrices and because some of the 
largest matrices and most important analysis tasks operate on graph datasets, 
e.g.~social networks and Web graphs. 
Facebook's social network has billions of vertices and
the Web Data Commons graph \cite{web_graph} has 3.6 billion vertices and 128
billions edges. We perform sparse matrix multiplication for graph
analysis such as community detection with NMF and spectral analysis with SVD.

%


Current research focuses on sparse matrix multiplication in memory
for small matrices and scaling to a large sparse matrix in a large cluster,
where the aggregate memory is sufficient to store the sparse matrix
\cite{Williams07, Yoo11, Boman2013}.
Distributed solutions for sparse matrix multiplication lead to significant
network communication and network bandwidth is usually the bottleneck.
As such, this operation requires a fast network to achieve performance.
A supercomputer or a large cluster with a fast network is inaccessible or
too expensive for many users. In addition, the distributed solution imposes
challenges in achieving load balancing on power-law graphs.

On the other hand, a current trend for hardware design scales up
a single machine, rather than scaling out to 
many networked machines.
These machines typically have multiple processors with many CPU cores and
a large amount of memory. They are also equipped with fast flash
memory such as solid-state drives (SSDs) to further extend memory capacity.
This conforms to the node design for supercomputers \cite{Ang14}.
Recent finding, from Frank McSherry \cite{McSherry15,McSherryBlog} and our
prior work \cite{FlashGraph},
show that the largest graph analytics tasks can be done on a small fraction of
the hardware, at less cost, as fast, and using less energy on a single shared-memory 
node, rather than a distributed compute engine.  Our findings in this paper reveal
that sparse matrices have the same structure and benefits and that the largest graph
matrices can be processed efficiently on a single scale-up compute node.

We explore a solution that scales sparse matrix dense matrix multiplication
(SpMM) on a multi-core machine with commodity SSDs and
perform SpMM in semi-external memory (SEM). The concept of semi-external memory
arose as a functional computing approach for graphs \cite{Abello98} in which
the vertex state of a graph is stored in memory and the edges accessed from
external memory. We introduce a similar construct for SpMM in which one or more
columns of a dense matrix are kept in memory and the sparse matrix is accessed
from external memory. In semi-external memory, we assume
that the memory of a machine is sufficient to keep at least one column
of the input dense matrix but is insufficient to hold the sparse matrix
and the dense matrices. Given fast SSDs, we demonstrate that the SEM
solution uses the resources of a multi-core machine well and
achieves performance that exceeds the state-of-the-art in-memory implementations.


We overcome many technical challenges to construct a sparse matrix
multiplication implementation on SSDs to achieve performance. Specifically,
SSDs are an order of magnitude slower in throughput and multiple orders of
magnitude slower in latency than DRAM. Furthermore, sequential I/O of SSDs
is still much faster
than random I/O \cite{safs} and reads are faster than writes. In addition,
SSDs wear out when we write data to them and random writes further shorten
the lives of SSDs \cite{sfs}. As such, our solution needs to sequentialize
I/O access and reduce I/O, especially writes.

Semi-external memory provides a scalable and efficient SpMM solution that
meets the I/O challenges.  We further incorporate substantial computation optimizations
to achieve performance of SpMM on graphs in a NUMA (non-uniform memory
architecture) machine. During the computation, each
thread streams its own partitions of the sparse matrix from SSDs, maximizing
I/O throughput and avoiding thread synchronization. We buffer all
intermediate computation results in local memory, to reduce remote memory
access, and stream the output matrix to SSDs at most once, minimizing writes
to SSDs. We design a very compact sparse matrix format to accelerate retrieving
a sparse matrix from SSDs. Semi-external memory has memory constraints.
As such, we deploy only computation optimizations that require a small memory
footprint, e.g., fine-grain dynamic task scheduling to balance loads on
power-law graphs and cache blocking to increase CPU cache hits.

Given the sparsity and the enormous size of graphs, the dense matrices involved
in SpMM are massive and our semi-external memory solution adapts to
dense matrices with different numbers of columns. When the dense matrix is
larger than memory, we split it
vertically into multiple partitions so that each partition can fit in
memory. As such, the minimum memory requirement of our solution is $O(r)$,
where $r$ is the number of rows in the input dense matrix. By keeping more columns
in the dense matrix in memory, we reduce I/O from SSDs and SEM-SpMM becomes
CPU bound, instead of I/O bound, on fast SSDs.

We develop three important applications in scientific computing and graph
analysis
with our SEM-SpMM: PageRank \cite{pagerank}, eigensolver \cite{anasazi} and
non-negative matrix factorization \cite{nmf}. Each of them requires SpMM with
different numbers of columns in dense matrices, resulting in different
strategies of placing data in memory. With the three applications, we
demonstrate data placement choices for different memory capacities in a machine
and the impact of the memory size on the performance of the applications.


The main contributions of this paper are:
\begin{itemize}
	\item We scale SpMM in semi-external memory using asymmetric I/O to SSDs and we
    deploy runtime optimizations that meet memory constraints.
	\item We deploy a compact sparse matrix format to reduce the amount of
		I/O and alleviate I/O as the bottleneck of the system.
	\item We deploy a fine-grain dynamic load balancing scheme for sparse
		matrices with power-law distribution in non-zero entries.
	\item We optimize SpMM for dense matrices with various numbers of columns
		and extends SEM-SpMM to large dense matrices that cannot fit in memory
		by vertically partitioning the dense matrices.
\end{itemize}

Our result shows that for real-world sparse graphs, our SEM-SpMM achieves almost
100\% performance of our in-memory implementation on a large parallel machine
with 48 CPU cores when the dense matrix has more than four columns. Even for
{\em SpMV} (sparse-matrix vector multiplication: the special case in which the dense 
matrix has a single column), our SEM implementation achieves at least 65\% 
performance of our in-memory implementation and outperforms Trilinos \cite{trilinos} and MKL \cite{mkl} by
a factor of $2-9$. 
We conclude that semi-external memory coupled with SSDs delivers an efficient
solution for large-scale sparse matrix multiplication. It serves
as a building block and offers new design possibilities for large-scale
data analysis, replacing memory with larger, cheaper, more energy-efficient SSDs
and processing bigger problems on fewer machines. The code for SpMM and SpMV has been 
released as open source as part of the FlashX system at http://flashx.io.

\section{Related Work}
Recent sparse matrix multiplication studies focus on in-memory optimizations.
Williams et al.~\cite{Williams07} describe optimizations for sparse matrix
vector multiplication (SpMV) in multicore architectures. Yoo et al.~\cite{Yoo11}
and Boman et al.~\cite{Boman2013} optimize distributed SpMV for large
scale-free graphs with 2D partitioning to reduce communication between
machines. In contrast, sparse matrix dense matrix multiplication (SpMM) receives
less attention from the high-performance computing community. Even though
SpMM can be implemented with SpMV, SpMV fails to explore data locality in
SpMM. Aktulga et al.~\cite{Aktulga14} optimize SpMM with cache blocking.
Koanantakool et al.~\cite{Koanantakool16} experiment with different parallel
algorithms for sparse matrix dense matrix multiplication and analyze
their communication cost in distributed memory.
We advance SpMM with a focus on optimizations for semi-external memory.

Compressed row storage (CSR) and compressed column storage (CSC) formats are commonly
used sparse matrix formats in many numeric libraries, such as Intel MKL \cite{mkl}
and Trilinos \cite{trilinos}. However, these formats are not designed for graphs.
Sparse matrix multiplication with these formats on graphs incurs many random memory
accesses. 
Sparsity \cite{Im04} designs a format that encodes both register blocking and cache blocking to
increase data reuse in the CPU cache for sparse matrix multiplication. Register blocking
requires explicit storage of zero values in register blocks. This strategy
wastes space and computation for graphs because graphs are sparse and
a vertex in a graph can connect to arbitrary other vertices. Buluc et al.~\cite{Buluc08}
further advance sparse matrix format
by doubly compressed sparse column (DCSC) for hypersparse submatrices after 2D
partitioning on a sparse matrix. This format significantly reduces the storage
size of a 2D-partitioned sparse matrix. Our format further reduces the storage
size of a sparse matrix.

Abello et al.~\cite{Abello98} introduced the semi-external memory algorithmic
framework for graphs. Pearce et al.~\cite{Pearce10} implement several 
semi-external memory graph traversal algorithms for SSDs. FlashGraph
\cite{FlashGraph} adopted the concept and performs graph algorithms with
vertex state in memory and edge lists on SSDs. This work extends the semi-external
memory concept to matrix operations.

GridGraph \cite{gridgraph} is a general-purpose graph processing framework on
a single machine
and scales to large graphs using disks. It performs 2D partitioning on a graph
to reduce CPU cache misses in graph analysis and deploys 2-level hierarchical
partitioning to reduce I/O. This graph engine is designed for graph
algorithms expressed as sparse matrix vector multiplication and keeps both
graphs and vertex computation state on disks. In contrast, our work focuses on
optimizations on sparse matrix dense matrix multiplication and performs this
operation in semi-external memory to reduce I/O, especially writes,
to SSDs. GridGraph runs more slowly than in-memory linear algebra libraries
such as Intel MKL \cite{mkl} and Trilinos Tpetra \cite{trilinos}.

Array databases such as SciDB \cite{scidb} and Rasdaman \cite{rasdaman} support
sparse matrix mutliplication. They perform matrix multiplication using CSR
formats \cite{SLACID}, similar to Intel MKL and Trilinos.  There has also been
preliminary research on accelerating matrix multiplication with GPUs \cite{Liu14},
showing that speedups are limited by I/O and setup costs and that benefits
are limited to compute dense operations.


MapReduce \cite{MapReduce} is also commonly used for processing large graphs.
PEGASUS \cite{pegasus} is a popular graph processing engine built on MapReduce
and expresses graph algorithms as a generalized form of sparse matrix-vector
multiplication. GBase \cite{gbase} is a graph database built on MapReduce.
It optimizes the graph storage for graph queries and analysis and perform graph
queries and analysis with sparse matrix vector multiplication.
Other work \cite{Liao14, Yin14, Liu10} implement non-negative matrix
factorization (NMF) with MapReduce. Although these MapReduce-based
implementations can scale to very large graphs, they run orders of
magnitude more slowly than optimized in-memory solutions, such as Trilinos and Intel MKL.

Zhou et al.~\cite{Zhou12} implement an LOBPCG \cite{Arbenz05} eigensolver in
an SSD cluster. Their implementation targets nuclear many-body Hamiltonian
matrices, which are much denser and have smaller dimensions than many sparse
graphs. Therefore, their solution stores the sparse matrix on SSDs and keep
the entire vector subspace in RAM. In contrast, our SEM-SpMM handles dense
matrices of different sizes and our eigensolver stores both the sparse matrix
and the vector subspace on SSDs.

The Trilinos project \cite{trilinos} has a large collection of numerical libraries.
The Tpetra library provides efficient sparse matrix operations such as sparse matrix
multiplication. The matrix implementations in Trilinos are optimized for
distributed memory.

Intel Math Kernel Library \cite{mkl} is an efficient and parallel linear
algebra library with matrix operations optimized for Intel
platforms. It provides an efficient sparse matrix multiplication
for regular sparse matrices. In contrast, our sparse matrix multiplication
optimizes for power-law graphs with near-random vertex connection.

\section{Sparse matrix multiplication} \label{sec:spmm}
Sparse matrix dense matrix multiplication (SpMM) generates many random memory
accesses and its performance is limited by random memory throughput
of DRAM. We perform SpMM in semi-external memory (SEM) to enable
nearly in-memory performance, while achieving scalability in proportion
to the ratio of non-zero entries to rows or columns in a sparse matrix.

\subsection{Semi-external memory}
Our definition of semi-external memory for SpMM keeps the sparse matrix on
SSDs and the input dense matrix or some columns of the input dense matrix
in memory. We may keep the entire output matrix in memory if it is sufficiently
small or stream it to SSDs or the subsequent computation.

In some applications, such as non-negative matrix factorization (Section
\ref{sec:spmm:apps}), even the input dense matrix does not fit in memory. In this case,
we partition the input dense matrix vertically so that each partition has
complete columns of the original input dense matrix that fit in memory.
For each vertical partition, we perform SpMM in semi-external memory and
stream the output matrix to SSDs.

\subsection{Sparse matrix format}
We design a new compact sparse matrix format that increases CPU cache hits and 
reduces the amount of data read from SSDs. A compact format is performance critical 
because in SEM-SpMM the I/O to SSDs is most often the bottleneck.


\begin{figure}
\centering
\includegraphics[scale=0.3]{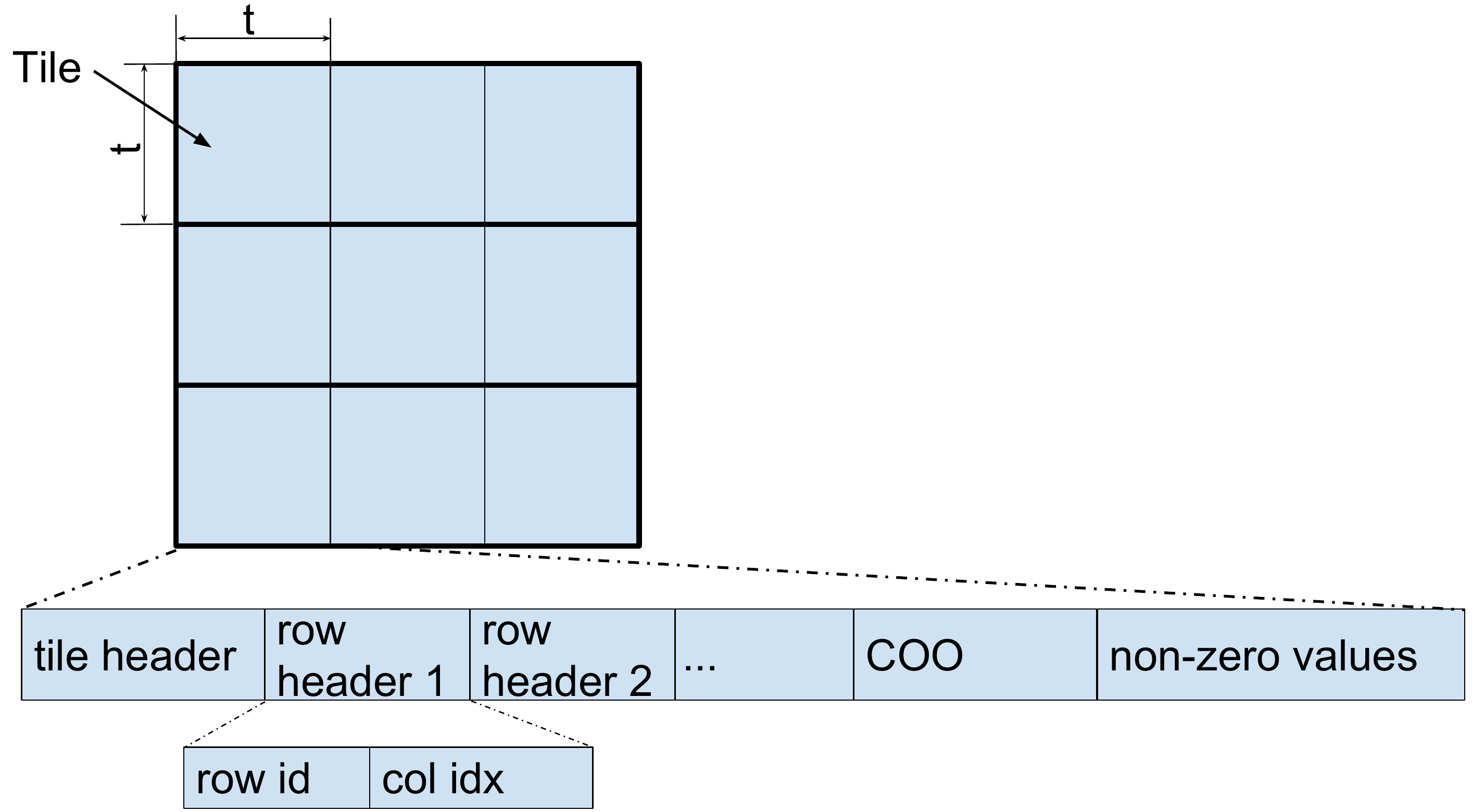}
\caption{The storage format of a sparse matrix. In this example, non-zero
entries of the sparse matrix are organized into 9 tiles. Tiles are stored in
the row-major order and a tile stores non-zero entries in SCSR+COO format.}
\label{sparse_mat}
\end{figure}

To increase CPU cache hits, we deploy cache blocking \cite{Im04} and store
non-zero entries of a sparse matrix in tiles, $t \times t$ submatrices (Figure
\ref{sparse_mat}).
When a tile is small, the rows from the input and output dense matrices
involved in multiplication with the tile are always kept in the CPU cache
during the computation. The optimal tile size should fill the CPU cache
with the rows from the dense matrices and is affected by the number of columns
of the dense matrices. To handle dense matrices with different numbers
of columns, we deploy both static cache blocking and dynamic cache blocking.
We generate sparse matrices with a relatively small tile size and
rely on the runtime system
to optimize for different numbers of columns (Section \ref{sec:exec}).
However, a small tile size potentially increases the storage size of a sparse
matrix. In semi-external memory, the dense matrices usually have
a small number of columns in sparse matrix multiplication. We
use the tile size of $16K \times 16K$ by default to balance the matrix storage
size and the adaptibility to different numbers of columns.


We use a compact format to store non-zero entries and refer to this
format as SCSR (Super Compressed Row Storage) (Figure \ref{sparse_mat}).
In sparse matrices encoding graphs, many rows in a tile
do not have any non-zero entries. The CSR or CSC formats waste space because
they require an entry for each row/column in the row/column index. Doubly
compressed sparse column (DCSC) \cite{Buluc08} is proposed to avoid this problem.
However, even DCSC wastes space because it requires the storage of
pointers to columns and fails to reference to non-zero values with small
integers (e.g., two-byte integers). SCSR is designed to further shrink the storage
size of a tile. This format keeps data only for rows with non-zero entries in
a tile. Each non-empty row has a row header that only contains an identifier
to indicate the row number, followed by column indices. To determine the end
of a row, the most significant bit of the identifier is always set to 1, while
the most significant bit of a column index entry is always set to 0.
Owing to the small size of a tile, we use two bytes to store a row
number and a column index entry, which further reduces the storage size.
Each non-zero entry requires at most four bytes to indicate its location in
a matrix. Because the most significant bit is used to indicate the beginning
of a row, this format supports a maximum tile size of $32K \times 32K$.

\begin{figure}
	\begin{center}
		\footnotesize
		\includegraphics[scale=1]{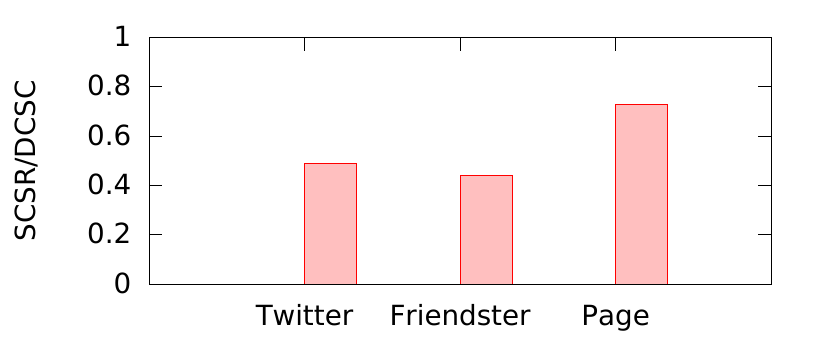}
		\caption{The ratio of the storage size required by SCSR and DCSC
		\cite{Buluc08} format for real-world graphs.}
		\label{fig:storage}
	\end{center}
\end{figure}

SCSR is much more compact than DCSC \cite{Buluc08}. 
Figure \ref{fig:storage} shows that SCSR uses 45\%-70\% of the storage size used by DCSC
for large real-world graphs (Table \ref{graphs}).
For a tile with $nnr$ non-empty rows and $nnz$ non-zero entries,
SCSR requires $2 \times nnr$ bytes for row ids, $2 \times nnz$ bytes for column
indices of non-zero entries and $c \times nnz$ bytes for non-zero values, where
$c$ is the number of
bytes for a non-zero value. As such, $S_{SCSR} = 2 \times nnr + (2 + c) \times nnz$.
In contrast, DCSC requires much more metadata for a tile
with $nnc$ non-empty columns and $nnz$ non-zero entries.
$S_{DCSC} = (2 + 2 + 4) \times nnc + (2 + c) \times nnz$. In a $t \times t$
tile, $nnr \le nnz \le nnr \times t$. On average, $nnr \approx nnc$ in most sparse
matrices. As such, for a binary sparse matrix, $0.4 \le S_{SCSR} / S_{DCSC} < 1$.
SCSR saves more space in a sparse matrix with larger $nnr / nnz$. 

Inside each cache tile of the SCSR, we use the coordinate format (COO) for
the rows that have only a single non-zero entry. For the adjacency matrices of
real-world graphs, many rows in a cache tile have only one non-zero entry.
Iterating over single-entry rows in the SCSR format requires to test
the end of a row for every non-zero entry, which leads to many conditional jumps.
In contrast, COO is more suitable for storing these
single-entry rows. It does not increase the storage size but significantly
reduces the number of conditional jump instructions. As a result, we combine
SCSR with COO and store non-zero entries in the COO format behind the row headers
of SCSR (Figure \ref{sparse_mat}).


\subsection{Dense matrices}
\begin{figure}
\centering
\includegraphics[scale=0.4]{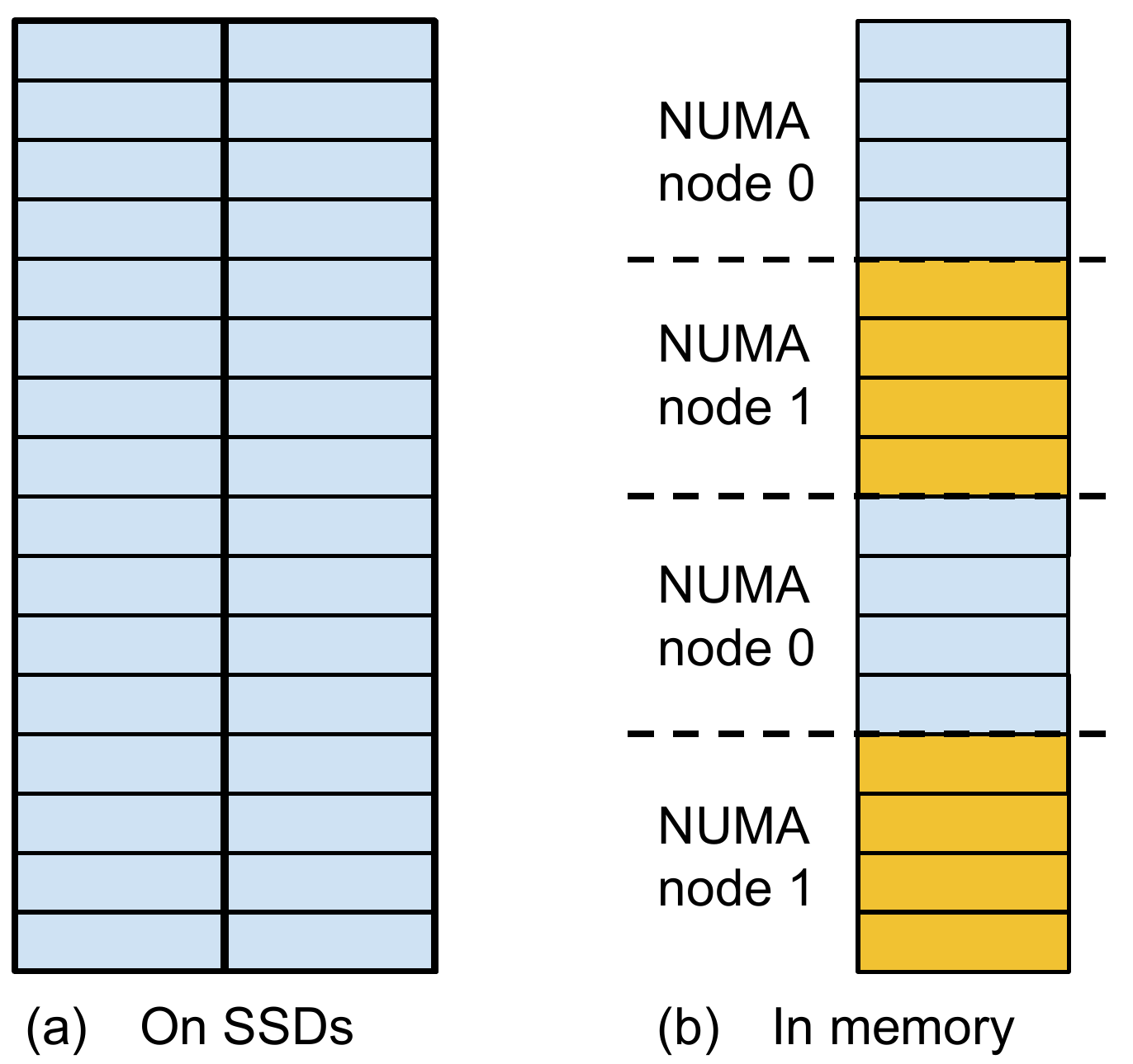}
\caption{Dense matrices are partitioned
vertically, so that each vertical partition fits in memory. A vertical
partition stores data in row-major order. When a vertical partition is loaded
into memory, it is partitioned horizontally into many row intervals that
are striped across NUMA nodes.}
\label{dense_mat}
\end{figure}

In many applications, the dense matrices in SpMM are tall and skinny with
millions or even billions of rows and a small number of columns.
Because many real-world graphs
are very large and sparse, the dense matrices may be too large to fit in memory.
Thus, for the general case of SpMM, we split the dense matrices into vertical
partitions of the memory size. To increase data locality in SpMM, the elements
in each vertical partition are stored in row-major order (Figure \ref{dense_mat}
(a)).  When performing SpMM in semi-external memory, we load the input dense
matrix or one of its vertical partitions into memory. 

For a non-uniform memory architecture (NUMA), we partition the input dense matrix
or one of its vertical partitions horizontally and store horizontal partitions
evenly across NUMA nodes (Figure \ref{dense_mat}(b)). Striping partitions evenly
across all NUMA nodes helps to fully utilize the bandwidth of memory.
We assign multiple contiguous rows in a row interval with $2^i$ rows to
a NUMA node. The row interval size is multiple of the tile size of
a sparse matrix so that multiplication on a tile accesses rows only
from a single row interval.

\subsection{Parallel Execution} \label{sec:exec}
\begin{figure}
\centering
\includegraphics[scale=0.3]{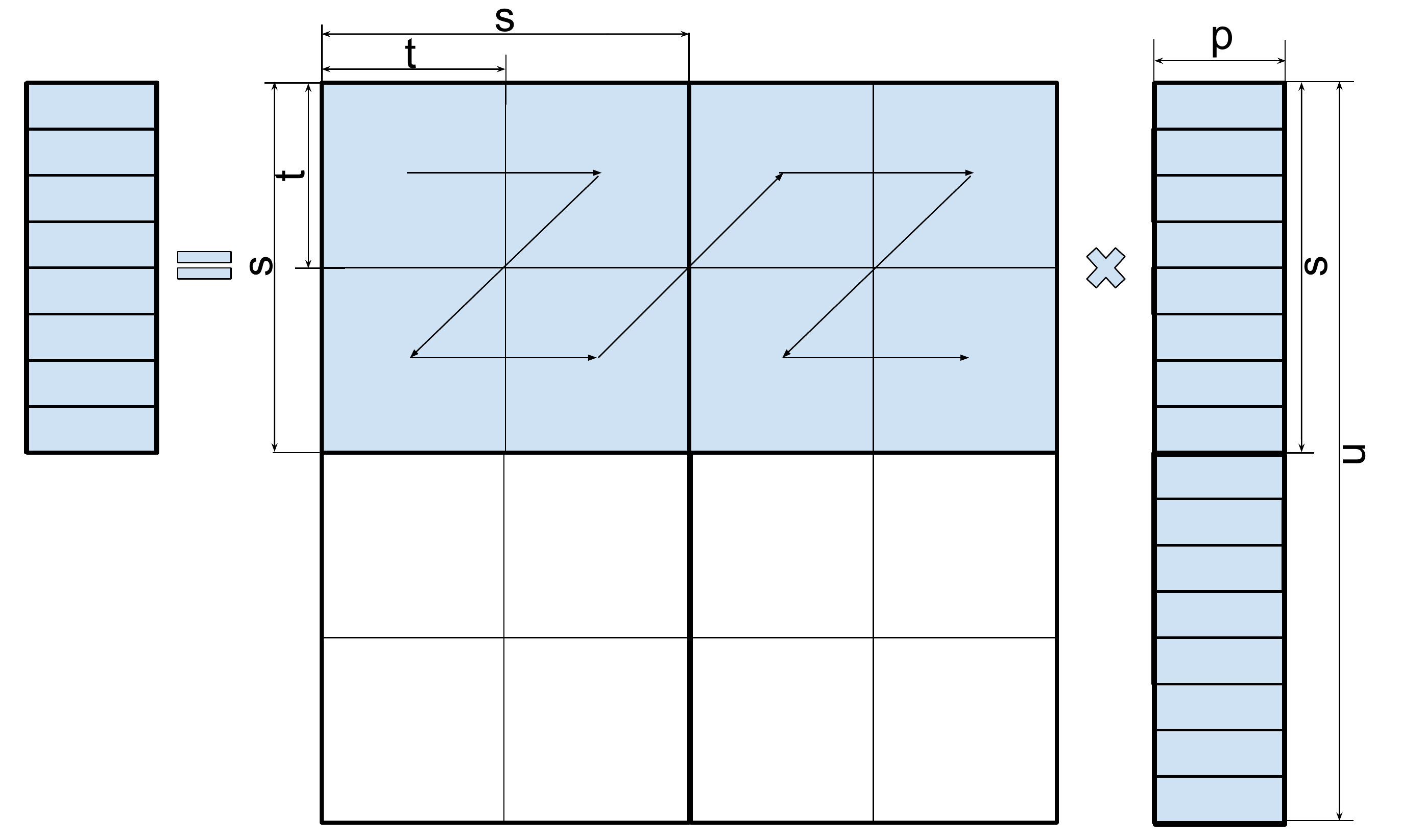}
\caption{The execution of multiplication on tiles in a sparse matrix and
	the input dense matrix in a single thread. A thread reads $\frac{s}{t}$
	tile rows and multiplies them with the input dense matrix.
	$s = \frac{CPU\_cache}{2 \times p}$ causes the dense matrices to fill
	the CPU cache. The arrows indicate the order of multiplication on tiles with the input dense matrix.}
\label{spmm_exec}
\end{figure}

We parallelize sparse matrix multiplication and deploy only computation
optimizations with a small memory footprint. Memory is precious resource
in semi-external memory because memory should be used to keep more columns
in the input dense matrix to reduce I/O from SSDs.
Algorithm \ref{alg:spmm} and Figure \ref{spmm_exec} illustrate the execution of
sparse matrix multiplication in semi-external memory.

Semi-external memory favors horizontal partitioning on a sparse matrix
for parallelization because this partitioning scheme minimizes writes to
remote memory and SSDs. Horizontal partitioning
requires only one thread to allocate local memory buffers for computation on
a tile row. All intermediate computation results on tiles are merged into
the local memory buffers. As such, we write the output matrix at most once
to SSDs and there are no remote memory writes.
In contrast, the vertical partitioning and 2D partitioning \cite{Koanantakool16}
would require each thread to maintain a local memory buffer for the same tile rows
to aggregate writes to SSDs and remote memory.

\begin{algorithm}
	\algblock{ParFor}{EndParFor}
	\algnewcommand\algorithmicparfor{\textbf{parfor}}
	\algnewcommand\algorithmicpardo{\textbf{do}}
	\algnewcommand\algorithmicendparfor{\textbf{end\ parfor}}
	\algrenewtext{ParFor}[1]{\algorithmicparfor\ #1\ \algorithmicpardo}
	\algrenewtext{EndParFor}{\algorithmicendparfor}

	\caption{Parallel execution of sparse matrix dense matrix multiplication.}
	\label{alg:spmm}
	\begin{algorithmic}[1]
		\Procedure{SparseMatrixMultiply}{$spm$, $inM$}
		\Statex \textbf{Input}: $spm$, a $n \times n$ sparse matrix on SSDs
		\Statex \textbf{Input}: $inM$, a $n \times p$ dense matrix in memory
		\Statex \textbf{Output}: $outM$, a $n \times p$ dense matrix on SSDs
		\State $\vec{trQ} \gets get\_tile\_row\_ids(spm)$
		\State $outM \gets zeros\_SSD(n, p)$
		\ParFor{$thread \in \vec{threads}$}
		\State $ProcessTileRows(spm, \vec{trQ}, inM, outM)$
		\EndParFor
		\EndProcedure

		\State

		\Procedure{ProcessTileRows}{$spm$, $\vec{trQ}$, $inM$, $outM$}
		\Statex \textbf{Input}: $spm$, a $n \times n$ sparse matrix on SSDs
		\Statex \textbf{Input}: $\vec{trQ}$, a queue that contains all tile row ids
		\Statex \textbf{Input}: $inM$, a $n \times p$ input dense matrix in memory
		\Statex \textbf{In-Out}: $outM$, a $n \times p$ output dense matrix on SSDs
		\State $t \gets get\_tile\_size(spm)$
		\State $numTRs \gets \frac{CPU\_cache\_size}{2 \times p \times t}$
		\While{$|\vec{trQ}| > 0$}
		\If {$|\vec{trQ}| <= \#threads$}
		\State $numTRs \gets 1$
		\EndIf
		\State $\vec{ids} \gets get\_tile\_rows(\vec{trQ}, numTRs)$
		\State $trs \gets read\_tilerow\_async(spm, \vec{ids})$
		\State $res \gets MulTRows(trs, inM)$ when $trs$ is ready
		\State $write\_rows\_async(outM, res)$
		\EndWhile
		\EndProcedure

		\State

		\Procedure{MulTRows}{$trs$, $inM$}
		\Statex \textbf{Input}: $trs$, a $s \times n$ submatrix in the sparse matrix
		\Statex \textbf{Input}: $inM$, $n \times p$ dense matrix
		\Statex \textbf{Output}: $outBuf$, a $s \times p$ dense matrix
		\State $outBuf \gets zeros(s, p)$
		\State $t \gets get\_tile\_size(trs)$
		\For{$k=0; k < \frac{n}{t}; k = k + \frac{s}{t}$}
		\For{$i=0; i < \frac{s}{t}; i = i + 1$}
		\For{$j = 0; j < \frac{s}{t}; j = j + 1$}
		\State $tile \gets trs[i, k + j]$
		\State $lInMat \gets$ rows from $inM$ for $tile$
		\State $lOutMat \gets$ rows from $outBuf$ for $tile$
		\State $lOutMat += tile * lInMat$
		\EndFor
		\EndFor
		\EndFor
		\EndProcedure
	\end{algorithmic}
\end{algorithm}

We deploy a fine-grain dynamic load balancing scheme for parallel computation
with a global task queue \textit{trQ} (Algorithm \ref{alg:spmm}). Each thread
gets one computation task (tile rows) at a time from \textit{trQ} with
\textit{get\_tile\_rows} and performs computation (\textit{MulTRows}).
At the beginning, a thread gets larger computation tasks that contains multiple
tile rows; as the computation approaches completion, a thread gets smaller
tasks that contain only one tile row. This design reduces concurrent access to
the global data structure while realizing good load balance.

The algorithm performs asynchronous I/O and merge writes from multiple threads
into larger ones. Large writes ensure sustainable write throughput to SSDs and
increase their duration. To assist in I/O merging, we use
\textit{get\_tile\_rows()} to control a global execution
order that ensures that all threads are processing contiguous tile rows and
the results from the tile rows are located closely on SSDs.
We write computation results from \textit{MulTRows} to SSDs
asynchronously with \textit{write\_rows\_async()}, which merges the writes
from different threads.

To better utilize the CPU cache, \textit{MulTRows}, 
reorganizes $t \times t$
tiles from multiple contiguous tile rows into $s \times s$ blocks, where
$s = \frac{CPU\_cache}{2 \times p}$ (Figure \ref{spmm_exec}). We process all
tiles in a $s \times s$ block before moving to the next block. The tile size
of a sparse matrix is relatively small. As such, this execution order helps to
reuse data in the CPU cache from the computation on the previous tile.
\textit{MulTRows} also maintains
a local memory buffer $outBuf$ to store the computation results,
which minimizes remote memory access. Once the computation in \textit{MulTRows}
is complete, $outBuf$ contains complete results.

We vectorize the computation on rows of dense matrices.
In SpMM, we multiply a non-zero entry from a tile with all elements in
a row of the input dense matrix and add the results to the corresponding row
of the output dense matrix.
We perform these operations with vector CPU instructions, such as
AVX \cite{avx}, to enable more efficient memory access and computation.

\subsection{I/O optimizations}
Semi-external memory sparse matrix multiplication streams a sparse matrix from
SSDs. When we access fast SSDs sequentially, the overhead of operating systems,
such as thread context switches and memory allocation, becomes noticeable.
We tackle these obstacles to maximize I/O throughput.

We issue asynchronous I/O and poll for I/O to avoid thread
context switches.
I/O polling prevents a thread from being switched out after it completes all available computation.
Without polling, when a thread issues an I/O request and waits for I/O completion,
the operating system switches the thread
out; the operating system reschedules the thread for execution once I/O is
complete. However, there is latency for thread rescheduling and the latency
from frequent rescheduling can cause noticeable performance degradation
on a high-speed SSD array.

When accessing a sparse matrix or a dense matrix from SSDs, we maintain a set of
memory buffers for I/O access to reduce the overhead of memory allocation.
We use large I/O to access matrices on SSDs to increase I/O throughput.
Large memory allocation is expensive because the operating
system usually allocates a large memory buffer with \textit{mmap()} and
populates the buffer with physical pages. Therefore, we keep
a set of memory buffers allocated previously and reuse them for new I/O requests.
For accessing a sparse matrix, tile rows usually have different sizes, so we resize
a previously allocated memory buffer if it is too small for a new I/O request.

\subsection{The impact of the memory size on I/O}
\label{sec:spmm:mem}
More memory reduces I/O in semi-external memory. The minimum memory requirement
for SEM-SpMM is $n c + t \epsilon$, where $n$ is the number of rows
of the input dense matrix, $c$ is the element size in bytes,
$t$ is the number of threads processing the sparse matrix
and $\epsilon$ is the buffer size for the sparse matrix and the output
dense matrix. When a machine does not have sufficient memory to keep the entire
input dense matrix in memory, we need multiple passes on the sparse matrix to
complete the computation. Reducing memory consumption is essential
to achieve performance in semi-external memory. By keeping more columns of
the input dense matrix in memory, we reduce the number of I/O passes.

Although we can cache part of a sparse matrix,
keeping more columns of the input dense matrix in memory saves more I/O than
using the same amount of memory for the sparse matrix. Assume the input
dense matrix has $n$ rows and $p$ columns. Again, $c$ is the element size
in bytes. The storage size of the sparse
matrix is $E$ bytes and the memory size is $M$ bytes. We further assume
we use $M'$ bytes to keep some columns of the dense matrices in memory
($M' < M$, ${n c p} \mod {M'} \equiv 0$)
and the remaining memory ($M - M'$) to cache the sparse matrix.
The amount of data in the sparse matrix read from SSDs is
\begin{equation*}
IO_{in} = \frac{n c p}{M'} [E - (M - M')]
\end{equation*}
Because $E > M$ in semi-external memory, we minimize $IO_{in}$ by maximizing $M'$,
i.e., using more memory for the input dense matrix.

As the number of columns in memory from the input dense matrix increases,
the bottleneck of the system
may switch. When we keep only one column of the input dense matrix in memory,
the system is usually I/O bound; when we keep more columns of the dense matrix
in memory, the system will become CPU bound and the I/O complexity does not
affect its performance.

\section{Applications} \label{sec:spmm:apps}
We apply sparse matrix multiplication to three important applications widely
used in graph analysis and machine learning: PageRank \cite{pagerank},
eigensolver \cite{anasazi} and non-negative matrix factorization \cite{nmf}.
Each application demonstrates a different strategy of using memory for SpMM.

\subsection{PageRank} \label{sec:pagerank}
PageRank is an algorithm to rank the Web pages by using hyperlinks between Web
pages. It was first used by Google and is identified as one of the top 10 data
mining algorithms \cite{top10}. PageRank is a representative of a set of graph
algorithms that can be expressed with SpMM or generalized SpMM, in which
the dense matrices have only one column. Other important examples are label propagation
\cite{label_prop} and belief propagation \cite{Yedidia03}. The algorithm runs
iteratively and its update rule for each Web pages in an iteration is
\begin{equation*}
PR(u) = \frac{1-d}{N} + d \sum\limits_{v \in B(u)} \frac{PR(v)}{L(v)}
\end{equation*}
where $B(u)$ denotes the neighbor list of vertex $u$ and $L(v)$ denotes
the out-degree of vertex $v$. 



\subsection{Eigensolver}
An eigensolver is another commonly used application that requires sparse matrix
multiplication. Many algorithms \cite{Lanczos, IRLM, krylovschur} and frameworks
\cite{arpack, anasazi, slepc} have been developed to solve a large eigenvalue
problem.

We take advantage of the Anasazi eigensolver framework \cite{anasazi} and
replace its original matrix operations with our SEM-SpMM. To compute eigenvalues
of a $n \times n$ sparse matrix, many eigenvalue algorithms construct a vector
subspace with a sequence of sparse matrix multiplications and each vector in
the subspace has the length of $n$. Due to the sparsity of real-world graphs,
the vector subspace is large and we keep the vectors on SSDs.
The Anasazi eigensolvers have block extension to update multiple vectors in
the subspace simultaneously, which results in sparse matrix dense
matrix multiplication. The most efficient Anasazi eigensolver on sparse graphs
is the KrylovSchur eigensolver \cite{krylovschur}, which updates a small number
of vectors (1-4) in the subspace simultaneously. Zheng et al.
\cite{flasheigen} provides the details of extending the Anasazi eigensolver.


\subsection{Non-negative matrix factorization}
Non-negative matrix factorization (NMF) \cite{nmf} finds two non-negative
low-rank matrices $W$ and $H$ to approximate a matrix $A \approx WH$. NMF
has many applications in machine learning
and data mining. A well-known example is collaborative filtering \cite{cf} in
recommender systems. NMF can also be applied to graphs to find communities
\cite{yang13, wang11}.

Many algorithms are designed to solve NMF and here we describe an algorithm
\cite{nmf} that requires a sequence of sparse matrix dense matrix multiplications.
The algorithm uses multiplicative update rules and updates matrices $W$ and $H$
alternately. In each iteration, the algorithm first fixes $W$ to update $H$
and then fixes $H$ to update $W$.
\begin{equation*}
H_{a\mu} \leftarrow H_{a\mu} \frac{{(W^TA)}_{a\mu}}{{(W^TWH)}_{a\mu}},
W_{ia} \leftarrow W_{ia} \frac{{(AH^T)}_{ia}}{{(WHH^T)}_{ia}}
\end{equation*}

We apply SEM-SpMM to NMF differently
based on the memory size and the number of columns in $W$ and $H$. Due to
the sparsity of a graph, $W$ and $H$ may require storage as large as
the sparse matrix and can no longer fit in
memory. Therefore, we partition $W$ and $H$ vertically and run multiple
sparse matrix multiplications to compute $W^TA$ and $AH^T$, if the memory is not
large enough.


\section{Experimental Evaluation}

We evaluate the performance of our SEM-SpMM on multiple real-world billion-scale
graphs including a web-page graph with 3.4 billion vertices. We first measure
the performance of SEM-SpMM and compare it with our in-memory implementation
(IM-SpMM), which is simply the SEM-SpMM implementation with the sparse matrix
in memory. 
We also compare SEM-SpMM with state-of-the-art in-memory implementations in
Intel MKL (\textit{mkl\_dcsrmm}) and Trilinos Tpetra. We use Intel MKL 2015
and Trilinos v12.0.1 for the experiments. We demonstrate the effectiveness of
CPU and I/O optimizations on SEM-SpMM and evaluate the overall performance
of the applications in Section \ref{sec:spmm:apps}.

We conduct experiments on a NUMA machine with
four Intel Xeon E7-4860 processors, clocked at 2.6 GHz, and 1TB memory of
DDR3-1600. Each processor has 12 cores. The machine has three LSI SAS 9300-8e
host bus adapters (HBA) connected to a SuperMicro storage chassis, in which
24 OCZ Intrepid 3000 SSDs are installed. The 24 SSDs together are capable of
delivering 12 GB/s for read and 10 GB/s for write at maximum. The machine runs
Linux kernel v3.13.0. We use 48 threads for all implementations.

\begin{table}
\begin{center}
\footnotesize
\begin{tabular}{|c|c|c|c|c|}
\hline
Graph datasets & \# Vertices & \# Edges & Directed \\
\hline
Twitter \cite{twitter} & $42$M & $1.5$B & Yes \\
\hline
Friendster \cite{friendster} & $65$M & $1.7$B & No \\
\hline
Page graph \cite{web_graph} & $3.4$B & $129$B & Yes \\
\hline
RMAT-40 \cite{rmat} & 100M & 3.7B & Yes \& No \\
\hline
RMAT-160 \cite{rmat} & 100M & 14B & Yes \& No \\
\hline
\end{tabular}
\normalsize
\end{center}
\caption{Graph data sets. We construct a directed and undirected version for
both RMAT-40 and RMAT-160.}
\label{graphs}
\end{table}

We use the adjacency matrices of the graphs in Table \ref{graphs} for performance
evaluation. The smallest graph we use has 42 million vertices and 1.5 billion
edges. The largest graph is the Page graph with 3.4 billion vertices and 129
billion edges. We generate two synthetic graphs with R-Mat \cite{rmat} to fill
the size gap between the smallest and largest graphs. We construct a directed and
undirected version for each of the synthetic graphs because some applications
in Section \ref{sec:spmm:apps} run on directed graphs and others run on undirected
graphs. The real-world datasets are publicly available and the synthetic
datasets are generated with the \textit{boost} RMAT generator\footnote{We use the 
parameters of $a=0.57$, $b=0.19$, $c=0.19$,
$d=0.05$.}. We always use the undirected version of the synthetic graphs for
the performance evaluation of sparse matrix multiplication.

\subsection{SEM-SpMM vs. IM-SpMM}

\begin{figure}
	\footnotesize
	\centering
	\begin{subfigure}[b]{0.5\textwidth}
		\centering
		\includegraphics[scale=1]{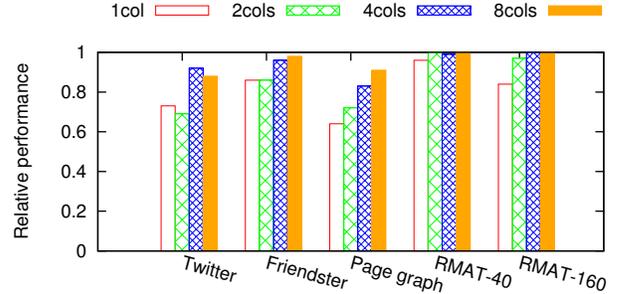}
		\vspace{-5pt}
		\caption{The runtime performance of SEM-SpMM, normalized to IM-SpMM
		for the dense matrix with the same number of columns.}
		\label{perf:spmm_comp}
	\end{subfigure}
	\begin{subfigure}[b]{0.5\textwidth}
		\centering
		\vspace{5pt}
		\includegraphics[scale=1]{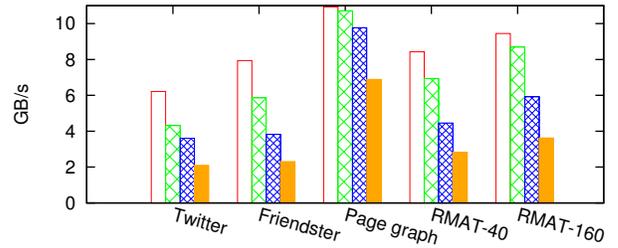}
		\vspace{-5pt}
		\caption{The I/O throughput generated by SEM-SpMM.}
		\label{perf:spmm_IO}
	\end{subfigure}
	\vspace{3pt}
	\caption{The performance and the I/O throughput of SEM-SpMM with dense
	matrices of different numbers of columns.}
	\label{perf:spmm_IM_vs_SEM}
\end{figure}

We compare the performance of SEM-SpMM with IM-SpMM to investigate
the performance penalty of scaling SpMM with SSDs. In this case, the dense
matrices have a small number of columns and are stored in memory. 

There is only a small performance penalty for semi-external memory (Figure
\ref{perf:spmm_comp}). The performance gap between IM-SpMM and SEM-SpMM
is affected by randomness of vertex connection. The gap is smaller if
vertex connection in a graph is more random. The Page graph is relatively
well clustered, so SpMM on this graph is less CPU-bound than others.
Even for the Page graph, SEM-SpMM gets 65\% performance of IM-SpMM.
The other factor of affecting the performance gap is the number of columns
in the dense matrices. The gap gets smaller as the number of columns in
the dense matrices increases.

We further measure the average I/O throughput generated by SEM-SpMM to indicate
the bottleneck of the system (Figure \ref{perf:spmm_IO}).   These experiments vary 
the number of columns in the dense matrix from 1 (Sparse Matrix Vector multiplication, SpMV) 
to 8.  SpMV on the Page
graph saturates the I/O bandwidth of SSDs and is clearly bottlenecked by I/O.
SpMV on other graphs (except Twitter) also generates high I/O throughput,
which consumes memory bandwidth and potentially interferes with random memory
access in SpMV. SEM-SpMV on the Twitter graph takes only about 0.5 second to
complete and,
thus, startup overhead has significant impact on the average I/O throughput.
As the number of columns in the dense matrix increases, SEM-SpMM becomes more
CPU-bound and requires a small number of columns to achieve close to 100\%
performance of IM-SpMM.

\begin{figure}
	\begin{center}
		\footnotesize
		\includegraphics[scale=1]{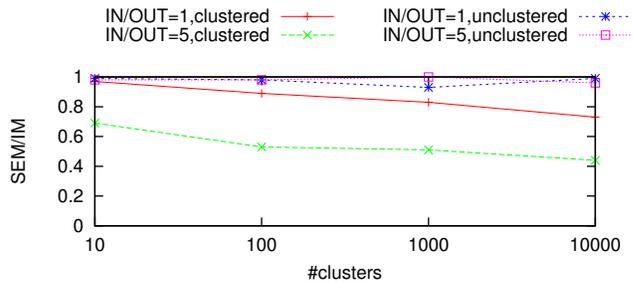}
		\caption{The relative performance of SEM-SpMV on graphs generated with
			stochastic block model, normalized to IM-SpMV on the same graphs.
			IN/OUT indicates the ratio of edges inside and outside clusters.
			A graph can have vertices ordered based on the cluster structures
		(clustered) and randomly ordered (unclustered).}
		\label{perf:spmm_sbm}
	\end{center}
\end{figure}

Multiple factors affect CPU cache hits in sparse matrix multiplication and
the performance gap between IM-SpMV and SEM-SpMV (Figure \ref{perf:spmm_sbm}).
We illustrate some of the factors with stochastic block model (SBM)
\cite{sussman12}, a random graph generation model with well-defined clusters
widely used in the graph community. In this experiment, we
generate SBM graphs with 100 million vertices and 3 billion edges.
If the vertices in a graph are randomly ordered,
SpMV on the graph generates many random memory access. Thus, the system is
bottlenecked by memory bandwidth and the performance gap between IM-SpMV and SEM-SpMV is small. 
On a graph with vertices ordered based on cluster structures, 
the number of clusters and
the ratio of edges inside and outside clusters have significant impact on
the performance gap.
As the number of clusters increases, the size of clusters gets smaller and
there are fewer random memory accesses inside clusters. Similarly, more
edges inside clusters also lead to fewer random memory accesses.
In both cases, the performance gap grows.

\subsection{Comparison with other in-memory SpMM}
In this section, we compare our SpMM implementation with the Intel MKL and
Trilinos Tpetra. Intel MKL runs on shared-memory machines. Trilinos Tpetra runs in
both shared memory and distributed memory, so we measure its performance in
our 48-core NUMA machine and an EC2 cluster. We run Tpetra in the largest
EC2 instances r3.8xlarge with 16 physical CPU cores and 244GB of RAM each.
The EC2 instances are connected with 10Gbps network in the same placement
group. MKL and Tpetra cannot run on the Page graph on our NUMA machine because
their memory consumption on this graph exceeds its memory capacity.

\begin{figure}
	\footnotesize
	\centering
	\begin{subfigure}[b]{0.5\textwidth}
		\centering
		\includegraphics[scale=1]{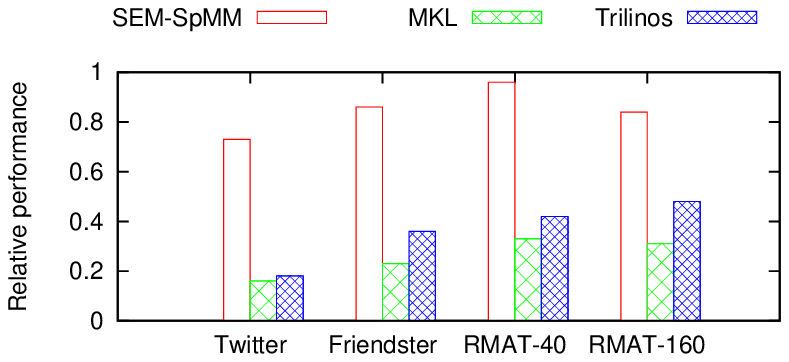}
		\vspace{-5pt}
		\caption{SpMV}
		\label{perf:spmv}
	\end{subfigure}
	\begin{subfigure}[b]{0.5\textwidth}
		\centering
		\includegraphics[scale=1]{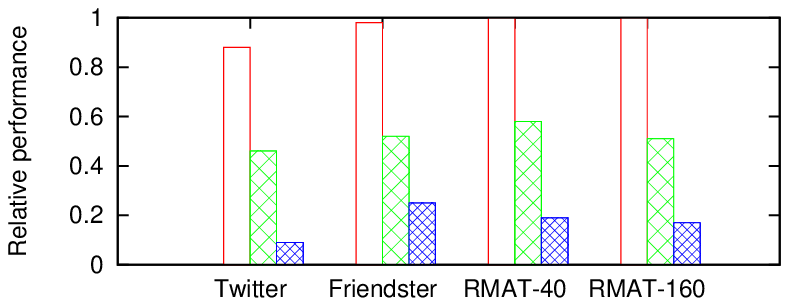}
		\vspace{-5pt}
		\caption{SpMM with a dense matrix of 8 columns.}
		\label{perf:spmm8}
	\end{subfigure}
	\vspace{3pt}
	\caption{The performance of different sparse matrix multiplication
		implementations on the 48-core machine normalized to IM-SpMM for
	the same graphs.}
	\label{perf:spmm}
\end{figure}

Our IM-SpMM and SEM-SpMM significantly outperform Intel MKL and
Trilinos Tpetra on the natural
graphs on our NUMA machine for both SpMV and SpMM (Figure \ref{perf:spmm}).
Even though the Tpetra implementation is optimized for SpMV, our SEM-SpMM
constantly outperforms Tpetra by a factor of $2-3$ even in SpMV. MKL has
better optimizations for SpMM than Tpetra. Our SEM-SpMM is still almost
twice as fast as MKL in SpMM with a dense matrix of eight columns. Our SpMM
implementation benefits all of the in-memory optimizations to achieve high
computation efficiency and has substantial I/O optimizations to reduce
the performance gap between IM-SpMM and SEM-SpMM (Section \ref{sec:opts}).
Specifically, the fine-grain dynamic load balancer achieves better load balancing
than MKL and Tpetra. Cache blocking significantly reduces CPU cache misses,
whereas MKL and Tpetra store a sparse matrix in CSC or CSR format and have
more CPU cache misses. 

\begin{figure}
	\begin{center}
		\footnotesize
		\includegraphics[scale=1]{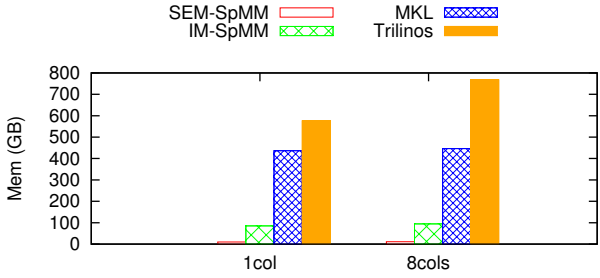}
		\caption{Memory consumption of different SpMM implementations on
		RMAT-160.}
		\label{perf:spmm_mem}
	\end{center}
\end{figure}

SEM-SpMM consumes a small fraction of memory when compared with IM-SpMM and
other SpMM implementations (Figure \ref{perf:spmm_mem}). SEM-SpMM consumes
memory for the input dense matrix and per-thread memory buffers for the sparse
matrix and the output dense matrix. The memory used by memory buffers
in each thread is significant
but is relatively constant for different graph sizes. We show the memory
consumption on one of the large graphs RMAT-160 (Figure \ref{perf:spmm_mem}).
SEM-SpMM uses about one tenth of the memory used by IM-SpMM, while IM-SpMM
consumes much less memory than MKL and Tpetra owing to its compact format for
sparse matrices.

\begin{figure}
	\footnotesize
	\centering
	\begin{subfigure}[b]{0.5\textwidth}
		\centering
		\includegraphics[scale=1]{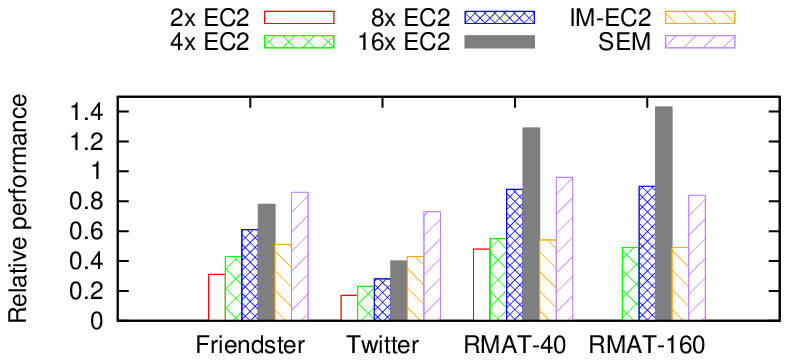}
		\vspace{-5pt}
		\caption{SpMV}
		\label{perf:ec2:spmv}
	\end{subfigure}
	\begin{subfigure}[b]{0.5\textwidth}
		\centering
		\includegraphics[scale=1]{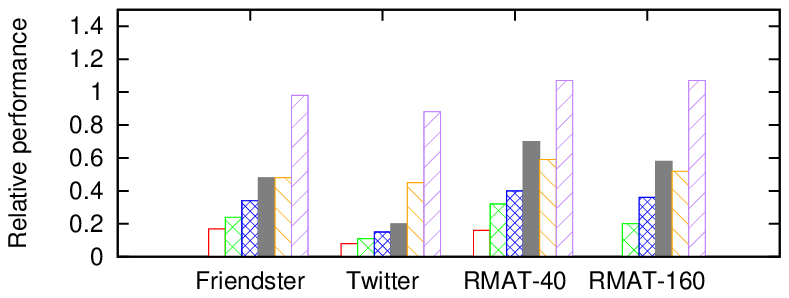}
		\vspace{-5pt}
		\caption{SpMM with a dense matrix of 8 columns.}
		\label{perf:ec2:spmm8}
	\end{subfigure}
	\vspace{3pt}
	\caption{The performance of SEM-SpMM on our 48-core machine (SEM) and
		Trilinos Tpetra on EC2 clusters (2xEC2, 4xEC2, 8xEC2 and 16xEC2),
		normalized to IM-SpMM on our 48-core machine for the same graphs.
		We also show the performance of IM-SpMM on
	one of the EC2 instance (IM-EC2) where Tpetra runs.}
	\label{perf:ec2}
\end{figure}

Our SEM-SpMM on a large parallel machine achieves comparable performance
and, in many cases, outperforms Trilinos Tpetra using five times
as much processing power, especially on
real-world graphs (Figure \ref{perf:ec2}). In this experiment, we run our
SpMM implementation on both our NUMA machine with 48 CPU cores
and one of the EC2 machines with 16 CPU cores. Owing to the compact format
for a sparse matrix, our IM-SpMM runs on all of the graphs on an EC2 instance
and achieves around half of the performance of our IM-SpMM on our NUMA machine.
When Tpetra runs on 16 EC2 instances, it has 5 times as many CPU cores as our
NUMA machine. Tpetra cannot run SpMV on RMAT-160 on two EC2 nodes.
Tpetra uses many more computation resources and still barely reaches
the same performance as our IM-SpMM and SEM-SpMM on our NUMA machine. One of
the main reasons that our SpMM implementation performs much
better on real-world graphs is that these graphs are more likely to cause
load imbalance. The dynamic scheduling of our SpMM implementation balances load much better than Tpetra.

\subsection{SEM-SpMM with a large input dense matrix}

We further measure the performance of SEM-SpMM with a large input dense matrix,
in which neither the sparse matrix nor the dense matrices can fit in memory.
In this experiment, we measure the performance of multiplying a sparse matrix
with a vertically partitioned dense matrix and the input and output dense matrices are
stored on SSDs. We study the impact of memory size on the performance of SEM-SpMM
by artificially varying the number of columns that can fit in memory. SEM-SpMM
accesses the sparse matrix with direct I/O and, thus, varying the number of
columns stored in memory does not affect I/O access to the sparse matrix.
IM-SpMM does not need to partition the dense matrices vertically.
We do not show the result on the Page graph because the dense matrix with
32 columns for this graph cannot fit in memory.

\begin{figure}
	\begin{center}
		\footnotesize
		\includegraphics[scale=1]{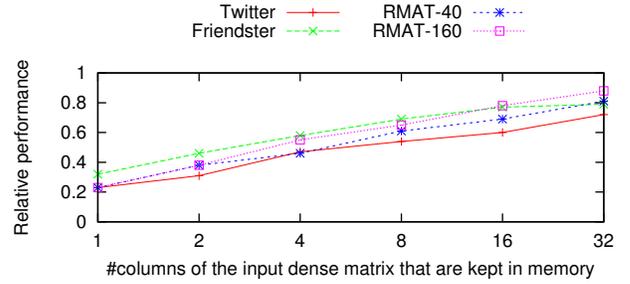}
		\caption{The performance of SEM-SpMM with a dense matrix of 32 columns
			relative to IM-SpMM, when the number of columns of the input dense
		matrix kept in memory varies.}
		\label{perf:spmm32}
	\end{center}
\end{figure}

As more columns in the input dense matrix can fit in memory, the performance
of SEM-SpMM constantly increases (Figure \ref{perf:spmm32}). When the memory
can fit over four columns of the input dense matrix, SEM-SpMM gets over 50\%
of the performance of IM-SpMM. Even when only one column of the input dense
matrix can fit in memory, SEM-SpMM still gets 25\% of the in-memory performance.
When the entire input dense matrix can fit in memory, we get about 80\% of
the in-memory performance.

\begin{figure}
	\begin{center}
		\footnotesize
		\includegraphics[scale=1]{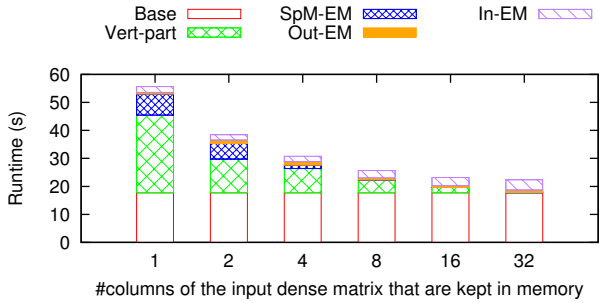}
		\caption{The overhead breakdown of SEM-SpMM on the Friendster
			graph with a dense matrix of 32 columns when the number
		of columns in the input dense matrix kept in memory varies. }
		\label{perf:spmm32_over}
	\end{center}
\end{figure}

Two main factors lead to performance loss in SEM-SpMM when the input dense matrix
cannot fit in memory. We illustrate the contribution of four potential overheads
in SEM-SpMM on the Friendster graph (Figure \ref{perf:spmm32_over}). The main
performance loss comes from the loss of data locality in SpMM caused by
vertical partitioning of the input dense matrix (Vert-part). Partitioning
the dense matrix into one-column matrices contributes 60\% of performance loss.
It drops quickly when the vertical
partition size increases. Keeping the sparse matrix on SSDs (SpM-EM)
also contributes some performance loss when the dense matrix is partitioned
into small matrices. The overhead almost goes away when more than four columns
of the dense matrix can fit in memory. The overhead of streaming the output dense
matrix to SSDs (Out-EM) and reading the input dense matrix to memory (In-EM)
is less significant and remains the same for different memory sizes.

\subsection{Optimizations on sparse matrix multiplication} \label{sec:opts}
Accelerating SEM-SpMM requires both computation and I/O optimizations.
We first evaluate the effectiveness of computation optimizations by deploying
them on IM-SpMM. We further show the effectiveness of I/O optimizations on
SEM-SpMM with all computation optimizations. The number of columns in
the dense matrix affects the effectiveness of the optimizations. As such,
we illustrate their effectiveness on SpMM for dense matrices with one column (SpMV)
up to eight columns.

Here we illustrate the most significant computation optimizations from Section
\ref{sec:spmm}. We start with an in-memory implementation that
performs sparse matrix multiplication on a sparse matrix in the CSR format
and apply the optimizations incrementally in the following order:
\begin{itemize} \itemsep1pt \parskip0pt \parsep0pt
	\item dispatch partitions of a sparse matrix to threads dynamically
		to balance load (\textit{Load balance}),
	\item partition dense matrices for NUMA (\textit{NUMA}),
	\item organize the non-zero entries in a sparse matrix into tiles to
		increase CPU cache hits (\textit{Cache blocking}),
	\item use CPU vectorization instructions to accelerate arithmetic
		computation (\textit{Vec}),
\end{itemize}

\begin{figure}
	\begin{center}
		\footnotesize
		\includegraphics[scale=1]{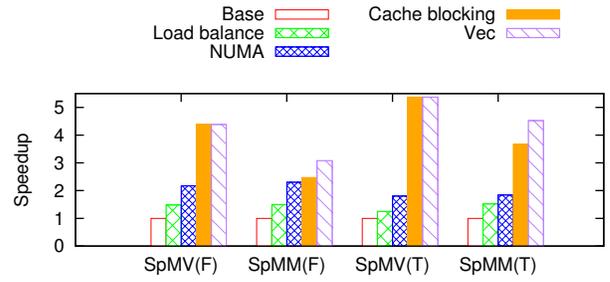}
		\caption{The speedup of each computation optimization for SpMV and SpMM
			on the Friendster (F) and Twitter (T) graph. The input dense matrix
		in SpMM has 8 columns.}
		\label{perf:spmm_opt}
	\end{center}
\end{figure}

All of these optimizations have positive effects on sparse matrix
multiplication and all optimizations together speed up SpMM by $3-5$ times
(Figure \ref{perf:spmm_opt}). The degree of effectiveness
varies between different graphs and different numbers of columns in
the dense matrices. The largest performance boost is from cache blocking,
especially for SpMV. This is expected because vertices in these graph can
connect to arbitrary vertices, which leads many random memory access in sparse matrix
multiplication. Cache blocking significantly increases CPU cache hits to reduce
random memory access. CPU vectorization is only effective on SpMM because
it optimizes computation on a row of the dense matrix.
With all optimizations, we have a fast in-memory implementation for both
sparse matrix vector multiplication and sparse matrix dense matrix multiplication.

We evaluate I/O optimizations on SEM-SpMV against a base implementation that
has all of the computation optimizations and use doubly compressed sparse row
format (DCSR) to store tiles of a sparse matrix. We illustrate their
effectiveness on the Friendster graph and the Page graph. The first one
represents a graph that is not well clustered; the other one is clustered with
domain names. We apply the I/O optimizations in the following order:
\begin{itemize} \itemsep1pt \parskip0pt \parsep0pt
	\item use SCSR to reduce data read from SSDs (SCSR),
	\item reduce memory allocation overhead for I/O with per-thread buffer
		pools (\textit{buf-pool}),
	\item reduce the number of thread context switches for I/O accesses with I/O
		polling (\textit{IO-poll}),
\end{itemize}

\begin{figure}
	\begin{center}
		\footnotesize
		\includegraphics[scale=1]{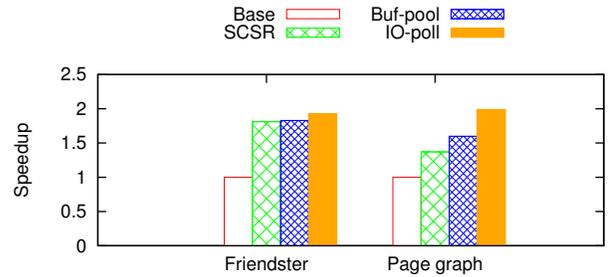}
		\caption{The speedup of I/O optimizations for SpMV on the Friendster
		graph and the Page graph.}
		\label{perf:spmm_opt_io}
	\end{center}
\end{figure}

The I/O optimizations lead to substantial speedup over the base implementation,
but behave very differently on these two graphs (Figure \ref{perf:spmm_opt_io}).
On the unclustered graph (Friendster), SCSR requires a much smaller storage size
than DCSR (Figure \ref{fig:storage}) and thus achieves significant
speedup. The Page graph, on the other hand, is well clustered and DCSR already
achieves a small storage size. SCSR further reduces the storage size, but is
less significant. SpMV on the Page graph has less random memory access and is
I/O-bound even on a large SSD array. \textit{Buf-pool} and \textit{IO-poll}
increases I/O throughput and, thus, improves performance. In contrast, SEM-SpMV
with the Friendster graph in the SCSR format already achieves almost 80\% of
IM-SpMV and, thus, further I/O optimizations have less noticeable speedup.

\begin{table}
\begin{center}
\footnotesize
\begin{tabular}{|c|c|c|c|c|}
\hline
	& Conv & Conv (I/O) & SpMV & SpMV (I/O) \\
\hline
Page graph & 86.00 s & 9.33 GB/s & 26.92 s & 10.60 s\\
\hline
RMAT-160 & 19.81 s & 9.51 GB/s & 7.96 s & 10.27 s\\
\hline
\end{tabular}
\normalsize
\end{center}
\caption{The speed and average I/O throughput of format conversion from CSR
	to SCSR on large graphs, compared with SEM-SpMV.}
\label{convert}
\end{table}

Format conversion from CSR to SCSR has low cost with linear time complexity and
is bottlenecked by SSDs on our largest graphs (Table \ref{convert}).
It reads the CSR image once and write the SCSR image back once, both sequentially.
This results in the minimum amount of I/O and I/O bounds the computation.
Format conversion generates only one-time overhead. Most of the applications
such as PageRank and eigensolving require many iterations of sparse matrix
multiplication to compute accurate results. As such, the format conversion
overhead is amortized.

\subsection{Application performance}

We evaluate the performance of our implementations of the applications in
Section \ref{sec:spmm:apps}. We show the effectiveness of additional memory for
these applications and compare their performance with state-of-the-art
implementations.

\subsubsection{PageRank}
We evaluate the performance of our SpMM-based PageRank implementation
(SpMM-PageRank). This implementation requires the input vector to be in memory,
but it is optional to keep the output vector and the degree vector in memory.
PageRank is a benchmarking graph algorithm implemented by many graph processing
frameworks. We compare the performance of SpMM-PageRank with state-of-the-art
implementations in FlashGraph \cite{FlashGraph}, a semi-external memory graph
engine, and GraphLab Create, the next generation of PowerGraph \cite{powergraph}.
The PageRank implementation in FlashGraph computes
approximate PageRank values while SpMM-PageRank and GraphLab Create compute
exact PageRank values. We run GraphLab Create completely in memory and
FlashGraph in semi-external memory. GraphLab Create is not able to compute
PageRank on the Page graph. We use FlashGraph v0.3 and a trial version of
GraphLab Create v1.9.

\begin{figure}
	\begin{center}
		\footnotesize
		\includegraphics[scale=1]{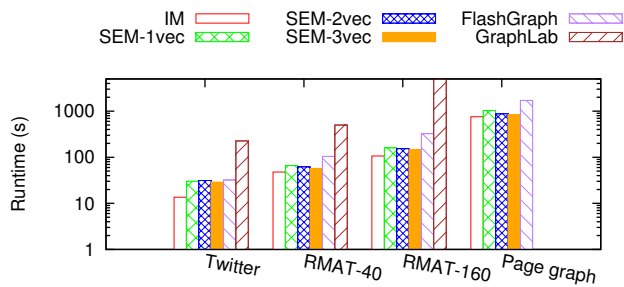}
		\caption{The runtime of SpMM-PageRank in 30 iterations. The SEM
			implementation keeps different numbers of vectors in memory
			(SEM-1vec, SEC-2vec, SEM-3vec). We compare them with
		the implementations in FlashGraph and GraphLab Create.}
		\label{perf:pagerank}
	\end{center}
\end{figure}

SpMM-PageRank in memory and in semi-external memory both significantly outperform
the implementations in FlashGraph and GraphLab Create (Figure \ref{perf:pagerank}).
Our SpMM is highly optimized for both CPU
and I/O. Even though SpMM-PageRank performs more computation than FlashGraph,
it performs the computation much more efficiently and
reads less data from SSDs than FlashGraph. SpMM-PageRank and the implementation
in GraphLab create performs the same computation, but SpMM-PageRank
performs the computation much more efficiently.

The experiment results also show that keeping more vectors in memory has modest
performance improvement for SpMM-PageRank. As such, SpMM-PageRank only needs
to keep one vector in memory, which results in very small memory consumption.

\subsubsection{Eigensolver}

We evaluate the performance of our SEM KrylovSchur eigensolver and compare
its performance
with our in-memory eigensolver and the Trilinos KrylovSchur eigensolver.
Usually, spectral analysis only requires a very small number of
eigenvalues, so we compute eight eigenvalues in this experiment. We run
the eigensolvers on the smaller undirected graphs
in Table \ref{graphs}. To evaluate the scalability of the SEM eigensolver,
we compute singular value decomposition (SVD) on the Page graph. Among all of
the eigensolvers, only our SEM eigensolver is able to compute eigenvalues
on the Page graph.

\begin{figure}
	\begin{center}
		\footnotesize
		\includegraphics[scale=1]{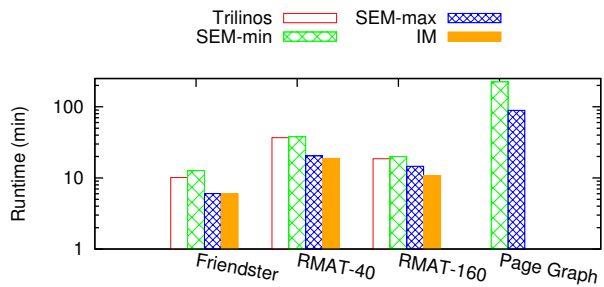}
		\caption{The runtime of our SEM KrylovSchur, our in-memory eigensolver
			and the Trilinos eigensolvers when computing eight
			eigenvalues. SEM-min keeps the entire vector subspace on SSDs and
		SEM-max keeps the entire vector subspace in memory.}
		\label{fig:eigen}
	\end{center}
\end{figure}

For computing 8 eigenvalues, our SEM eigensolver achieves performance
comparable to our in-memory eigensolver and the Trilinos eigensolver
and can scale to very large graphs (Figure \ref{fig:eigen}).
Unlike PageRank, an eigensolver has many more vector or dense matrix operations.
As such, the memory size has noticeable impact on performance.
For the setting with the minimum memory consumption, it has at least 45\%
performance of our in-memory eigensolver; when keeping the entire subspace
in memory, it has almost the same performance as our in-memory eigensolver.

\subsubsection{NMF}
We evaluate the performance of our NMF implementation (SEM-NMF) on the directed
graphs in Table \ref{graphs}. The dense matrices for NMF can be as large as
the sparse matrix. As such, we experiment with the effect of the memory size on
the performance of SEM-NMF by varying the number of columns in memory from
the dense matrices. We also compare the performance of SEM-NMF with
a high-performance NMF implementation SmallK \cite{SmallK}, built on top of
the numeric library Elemental \cite{elemental}. We factorize
each of the graphs into two $n \times k$ non-negative dense matrices and
we use $k=16$ because $16$ is the largest $k$ that SmallK supports for
the graphs in Table \ref{graphs}. We use SmallK v1.6 and Elemental v0.85.

\begin{figure}
	\begin{center}
		\footnotesize
		\includegraphics[scale=1]{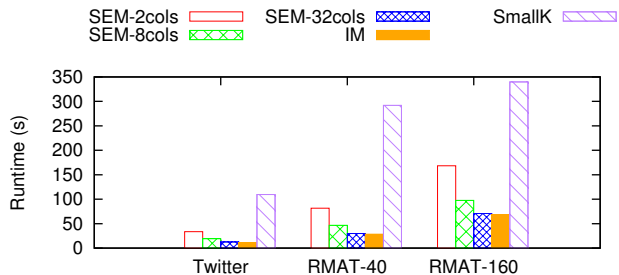}
		\caption{The runtime per iteration of SEM-NMF on directed graphs.
			We vary the number of columns in the dense matrices that are kept
			in memory to evaluate effect of the memory size on the performance
		of SEM-NMF.}
		\label{perf:NMF}
	\end{center}
\end{figure}

We significantly improve the performance of SEM-NMF by keeping more columns
of the input dense matrix in memory (Figure \ref{perf:NMF}). The performance
improvement is more significant when the number of columns that fit in memory
is small. When we keep eight columns of the input dense matrix in memory,
SEM-NMF achieves over 60\% of the performance of the in-memory implementation.

SEM-NMF significantly outperforms other NMF implementations.
SmallK is the closest competitor. We run the same NMF algorithm in SmallK and
SEM-NMF outperforms SmallK by a large factor on all graphs (Figure
\ref{perf:NMF}). 

\section{Conclusions}
We present an alternative solution for scaling sparse matrix dense matrix
multiplication (SpMM) to large sparse matrices by utilizing commodity SSDs
in a large parallel machine. We perform this operation in semi-external memory
(SEM), in which we keep the sparse matrix on SSDs and the dense matrices in
memory. Semi-external memory increases scalability in proportion to the ratio
of non-zero entries to rows or columns in a sparse matrix, while achieving
in-memory performance. We demonstrate
that our approach provides a promising alternative to distributed computation
for large-scale data analysis.

With substantial memory and I/O optimizations, our SEM-SpMM achieves high
efficiency while scaling to large graphs with
billions of vertices and hundreds of billions of edges. It significantly
outperforms the Intel MKL and Trilinos implementations. We run our
implementation in memory (IM-SpMM) to quantify the overhead of keeping data
on SSDs. SEM-SpMM achieves almost 100\% performance of IM-SpMM on some graphs
when the dense matrices have more than four columns, and achieves at least 65\%
of the performance of IM-SpMM on all graphs even when the dense matrix
has only one column.

For a machine with insufficient memory to keep the entire input dense matrix
in memory, we partition the dense matrix vertically and run SEM-SpMM multiple
times. In this case, the main overhead of SEM-SpMM comes from the loss of
data locality caused by vertical partitioning on the dense matrix. However,
given sufficient memory to keep a small number of columns of the input dense
matrix, we achieve performance comparable to IM-SpMM.

We apply our sparse matrix multiplication to three important applications:
PageRank, eigendecomposition and non-negative matrix factorization. We demonstrate
how additional memory should be used in semi-external memory in each application.
We further demonstrate that each of our implementations significantly outperforms
state of the art and scales to large graphs.

Our SSD-based solution also achieves high energy efficiency even though
we have not measured energy consumption explicitly. SSDs are energy-efficient
storage media \cite{Tsirogiannis} compared with RAM and hard drives.
When processing large datasets, our solution only uses
a single machine and requires a relatively small amount of memory. In contrast,
a distributed solution requires many more machines and much more aggregate
memory in order to process datasets of the same size. As such, our solution
introduces an energy-efficient architecture for large-scale data analysis tasks.


%

\ifCLASSOPTIONcompsoc
  \section*{Acknowledgments}
\else
  \section*{Acknowledgment}
\fi

This work is partially supported by NSF ACI-1261715,
DARPA GRAPHS N66001-14-1-4028 and
DARPA SIMPLEX program through SPAWAR contract N66001-15-C-4041.

\ifCLASSOPTIONcaptionsoff
  \newpage
\fi



\bibliographystyle{IEEEtran}
\bibliography{IEEEabrv,SpMM}
%




\begin{IEEEbiography}[{\includegraphics[width=1in,height=1.25in,clip,keepaspectratio]{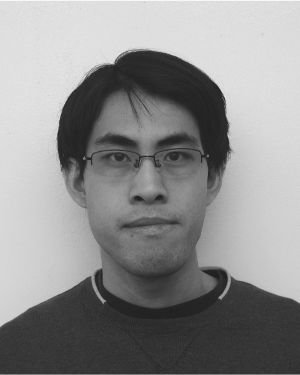}}]{Da Zheng}
	received the BS degree in computer science from Zhejiang University,
	China, in 2006, the MS degree in computer science from
	\'Ecole polytechnique f\'ed\'erale de Lausanne, in 2009. Since 2010,
	he is a PhD student of computer science at Johns Hopkins University.
	His research interests include high-performance computing, large-scale
	data analysis systems, programming languages and large-scale machine
	learning.
\end{IEEEbiography}

\begin{IEEEbiography}[{\includegraphics[width=1in,height=1.25in,clip,keepaspectratio]{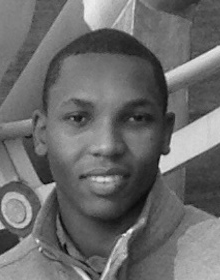}}]{Disa Mhembere}
	received the BS degree in electrical engineering from Morgan State University,
	in 2009, the MS degree in engineering management in 2012 and the MS degree
	in computer science in 2015 from Johns Hopkins University. He is a PhD student
	of computer science at Johns Hopkins University. His research interests include
	large-scale graph analysis and clustering.
\end{IEEEbiography}


\begin{IEEEbiography}[{\includegraphics[width=1in,height=1.25in,clip,keepaspectratio]{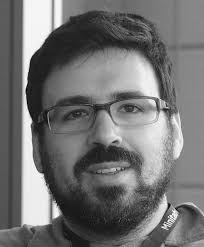}}]{Vince Lyzinski}
	received the BSc degree in mathematics from the University of Notre Dame
	in 2006, the MSc degree in mathematics from
	John Hopkins University (JHU) in 2007, the
	MScE degree in applied mathematics and statistics
	from JHU in 2010, and the PhD degree in
	applied mathematics and statistics from JHU in
	2013. From 2013 to 2014, he was a postdoctoral
	fellow in the Applied Mathematics and Statistics
	Department at JHU. Since 2014, he has
	been a senior research scientist at the JHU
	Human Language Technology Center of Excellence and an Assistant
	Research Professor in the AMS Department at JHU. His research interests
	include graph matching, statistical inference on random graphs,
	pattern recognition, dimensionality reduction, stochastic processes, and
	high-dimensional data analysis.
\end{IEEEbiography}


\begin{IEEEbiography}[{\includegraphics[width=1in,height=1.25in,clip,keepaspectratio]{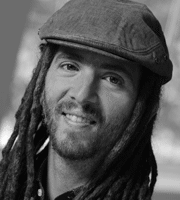}}]{Joshua T. Vogelstein}
	received the BS degree from the Department of Biomedical Engineering
	(BME) at Washington University in St. Louis,
	in 2002, the MS degree from the Department
	of Applied Mathematics and Statistics
	at Johns Hopkins University (JHU), in 2009, and the PhD degree
	from the Department of Neuroscience at JHU in
	2009. He was a postdoctoral fellow in AMS at
	JHU from 2009 until 2011, when he was
	appointed an assistant research scientist, and
	became a member of the Institute for Data Intensive Science and Engineering.
	He spent two years at Information Initiative at Duke, before
	becoming Assistant Professor in BME
	at JHU, and core faculty in the Institute for Computational Medicine
	and the Center for Imaging Science. His research interests include
	computational statistics, focusing on ultrahigh-dimensional
	and non-Euclidean neuroscience data, especially connectomics.
\end{IEEEbiography}

\vfill

\begin{IEEEbiography}[{\includegraphics[width=1in,height=1.25in,clip,keepaspectratio]{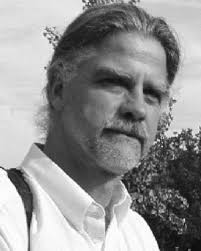}}]{Carey E. Priebe}
	received the BS degree in mathematics from Purdue University,
	in 1984, the MS degree in computer science from San Diego State
	University, in 1988, and the PhD degree in information technology
	(computational statistics) from George Mason University,
	in 1993. From 1985 to 1994, he worked as a
	mathematician and scientist in the US Navy
	research and development laboratory system.
	Since 1994, he has been a professor in the
	Department of Applied Mathematics and Statistics,
	Johns Hopkins University (JHU).
	His research interests include computational statistics, kernel and mixture
	estimates, statistical pattern recognition, statistical image analysis,
	dimensionality reduction, model selection, and statistical inference for
	high-dimensional and graph data. He is a lifetime member of the Institute
	of Mathematical Statistics, an elected member of the International Statistical
	Institute, and a fellow of the American Statistical Association.
\end{IEEEbiography}

\begin{IEEEbiography}[{\includegraphics[width=1in,height=1.25in,clip,keepaspectratio]{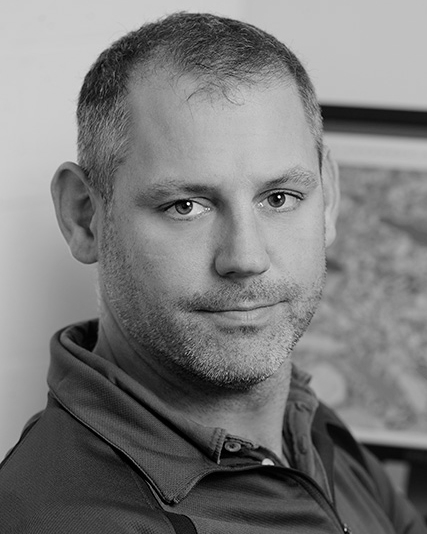}}]{Randal Burns}
	received the BS degree in Geophysics from Stanford University, in 1993,
	the MS degree in computer science from University of California, Santa Cruz,
	in 1997, and the PhD degree in in computer science from
	the University of California, Santa Cruz, in 2000. Since 2001,
	he has been been a professor in the Department
	of Computer Science at Johns Hopkins University,
	where he directs the Hopkins Storage Systems Lab, serves
	on the advisory board of Kavli Neuroscience Discovery Institute, is a memory of 
  the Institute for Data Intensive Engineering \& Science, and is a founder of NeuroData and the Open Connectome Project.
  His research focuses on the management
	and analysis of large datasets used in scientific applications.
\end{IEEEbiography}

\vfill

\end{document}